\documentclass[showpacs,floatfix,aps,prb,twocolumn]{revtex4}  
\usepackage{graphicx}
\usepackage{array}
\usepackage[usenames]{color}
\usepackage{amsmath}
\usepackage{amssymb}
\usepackage{ulem}
\newcommand{\tens}[1]{\ensuremath{{\bf #1}}}

\newcommand{\id}{\tens{1}}
\newcommand{\I}{\mathrm{i}}
\newcommand{\Tr}{\ensuremath{\operatorname{Tr}}}
\renewcommand{\Re}{\ensuremath{\operatorname{Re}}}

\newcommand{\emme}[1]{\tens{\bf M}_{#1}}

\newcommand{\red}[1]{{\color{black}{#1}}}
\newcommand{\evmax}{\Delta\epsilon}
\newcommand{\evmin}{\delta\epsilon}
\newcommand{\q}{l}

\begin{document}

\title{An efficient and accurate decomposition of the Fermi operator}
\author{Michele Ceriotti}\email{michele.ceriotti@phys.chem.ethz.ch}
\author{Thomas K\"uhne}
\author{Michele Parrinello}
\affiliation{Computational Science, Department of Chemistry and Applied Biosciences,
ETH Zurich, USI Campus, Via Giuseppe Buffi 13, CH-6900 Lugano, Switzerland}
\begin{abstract}
We present a method to compute the Fermi function of the Hamiltonian for a system
of independent fermions, based on an exact decomposition of the grand-canonical potential.
This scheme does not rely on the localization of the orbitals and is insensitive 
to ill-conditioned Hamiltonians. It lends itself naturally to linear scaling,
as soon as the sparsity of the system's density matrix is exploited.
By using a combination of polynomial expansion and Newton-like iterative techniques, 
an arbitrarily large number of terms can be employed in the expansion, 
overcoming some of the difficulties encountered in previous papers. 
Moreover, this hybrid approach allows us to obtain a very favorable scaling
of the computational cost with increasing inverse temperature, which makes the 
method competitive with other Fermi operator expansion techniques.
After performing an in-depth theoretical analysis
of computational cost and accuracy, we test our approach on the DFT Hamiltonian 
for the metallic phase of the $\mathrm{LiAl}$ alloy.
\end{abstract}

\pacs{
71.15.-m,
31.15.-p
}
\maketitle

\section{Introduction}
When calculating the various ground state properties of fermionic systems, it is important
to have fast and accurate ways of evaluating the density matrix. For non-interacting fermions,
this amounts to calculating the Fermi function associated with the system's Hamiltonian $\tens{H}$. 
In many applications, $\tens{H}$ is either of empirical nature, or the result of a self-consistent
\red{density functional theory (DFT)} calculation.
The standard method for computing the density matrix requires diagonalizing $\tens{H}$, 
an operation whose computational complexity scales cubically with the
number of electronic degrees of freedom $N$. 
Having a linear scaling scheme to obtain this quantity is a key step for 
modeling larger systems, thus making possible the computational study of a vast
class of problems, whose behavior cannot be described by smaller models. Areas
in which such a technique would have a major impact include nanotechnology and 
biochemistry, to name but a couple.

Several methods have been proposed to circumvent diagonalization\cite{goed99rmp}. 
These methods are based on the nearsightedness principle\cite{kohn96prl,prod-kohn05pnas},
which guarantees that in the $N\rightarrow\infty$ limit the matrices needed to 
compute the Fermi operator will become sparse. Among the different approaches that have
been proposed, we might cite divide-and-conquer schemes\cite{yang91prl},
density-matrix minimization\cite{li+93prb}, Green's function\cite{baro-gian92epl}, 
maximally localized orbitals\cite{gall-parr92prl} and 
penalty functions methods\cite{kohn96prl}. The use of sparse matrix algebra  
eventually leads to linear scaling\red{, both in terms of memory requirements and of 
computational cost}. 
A second class of methods, on which we shall focus here, uses the finite-temperature Fermi
operator. Due to the finite temperature, the singularity at the chemical potential $\mu$
is smoothed, thus allowing for an expansion in simpler functions of $\tens{H}$.
Since orbital localization is not explicitly exploited, this class of methods can also be 
applied to metals.
The earliest attempts in this direction were based on an expansion in 
Chebyshev polynomials\cite{goed-colo94prl,goed-tete95prb}. The computational cost
of this method has been analyzed by Baer and Head-Gordon\cite{baer-head97jcp}, 
who found that the order $m$ of the polynomial needed to achieve a $10^{-D}$ accuracy
depends linearly on the width of the  Hamiltonian spectrum $\Delta E$ and 
the electronic temperature $1/\beta$, i.e. $m\sim D \beta \Delta E$. 
This obviously raises some problems when considering Hamiltonians 
with large $\Delta E$, such as those arising from DFT calculations using plane wave
basis sets, or when low temperatures are required.
Recently it has been suggested\cite{lian+03jcp} that fast polynomial summation methods, requiring a
number of multiplications $\sim\sqrt{m}$, can be applied to Fermi operator expansion, 
leading to the more favorable scaling $\sqrt{\beta \Delta E}$.

In this paper we revisit a particular form for the expansion of the Fermi operator, 
which is based on the grand-canonical formalism and developed in a series of recent 
papers\cite{kraj-parr05prb,kraj-parr06prb,kraj-parr06prb-2,kraj-parr07prb}.
The grand-canonical potential for independent fermions is split into 
a sum of $P$ terms, containing $e^{-\beta\left(\tens{H}-\mu\right)/2P}$. 
As a consequence of this decomposition, the Fermi operator can be written
exactly as a sum of $P$ terms.
The larger the number of terms, the easier the evaluation of the exponential:
this implies a tradeoff between the size of $P$ and the accuracy of the results.
In this paper, we investigate the analytical properties of this decomposition, finding
that a large number of terms are almost ideally conditioned, and that their 
contribution to the Fermi operator can be easily and effectively computed
in a single shot with a polynomial expansion. The remaining few are tackled via a 
Newton-like iterative inversion scheme, which needs to be applied to each term individually
but is very efficient in dealing with large $D\beta\Delta E$.
With this hybrid approach, large values of $P$ can be reached at a cost
that is modest and independent of the system size. This result 
can improve significantly the prefactor of other methods using similar 
decompositions\cite{kraj-parr05prb,kraj-parr06prb,kraj-parr06prb-2,kraj-parr07prb}.
Moreover, using this approach, we achieve a scaling of the operations count with $D\beta\Delta E$
that is sublinear, and competitive with the result of Ref.\cite{lian+03jcp}
if their fast summation technique is used. In this way, accurate, low-temperature calculations
can be performed.

\section{Properties of the expansion\label{sec:foe-properties}}
We use an expansion of the Fermi operator based on grand-canonical 
formalism, which has been developed and employed in several recent 
works\cite{kraj-parr05prb,kraj-parr06prb,kraj-parr06prb-2,kraj-parr07prb}. 
We summarize the derivation and the resulting expression here, 
introducing a slightly different notation. To simplify the expressions, we will set the zero of 
energy at $\mu$, and measure energies in units of $k_B T$. This amounts to replacing
in the standard expression for the Fermi operator $\beta\left(\tens{H}-\mu \id\right)$
with $\tens{H}$.
Using this notation, the grand-canonical potential for a system of 
non-interacting fermions becomes\cite{alav-frenk92jcp,alav+94prl}
\begin{equation}
\Omega=-2\ln\det\left(\id+e^{-\tens{H}}\right)=
	-2\Tr\ln\left(\id+e^{-\tens{H}}\right).
\label{eq:grand-canonical}
\end{equation}
Introducing the $\emme{\q}$ matrices,
\begin{equation}
\emme{\q}=\id-e^{\I\left(2\q-1\right)\pi/2P}e^{-\tens{H}/2P}\label{eq:emme-q}
\end{equation}
we can perform the decomposition
\begin{equation}
 \left(\id+e^{-\tens{H}}\right)=\prod_{\q=1}^{P}\emme{\q}\emme{\q}^{\star}.\label{eq:mq-factor}
\end{equation}
These expressions are analogous to those introduced in Ref.\cite{kraj-parr05prb}, 
apart from a change of indices ($P/2\rightarrow P$, $P+1/2-l\rightarrow \q$).

Using factorization~(\ref{eq:mq-factor}), the grand-canonical potential can
be written in compact form as 
$\Omega=-2\Tr\sum_{\q=1}^{P}\ln\left(\emme{\q}\emme{\q}^{\star}\right)$.
The observables of interest for the system can be obtained as derivatives of 
the grand-canonical potential. In particular, the grand-canonical density matrix reads:
\begin{equation}
\tens{\boldsymbol{\rho}}=\frac{\delta\Omega}{\delta\tens{H}}=
\frac{1}{1+e^{\tens{H}}}=
\frac{2}{P}\sum_{\q=1}^{P}\id-\Tr\Re\emme{\q}^{-1}.\label{eq:density-matrix}
\end{equation}

The decomposition~(\ref{eq:density-matrix}) is exact for any value of $P$. As $P$ increases,
 the exponential $e^{-\tens{H}/2P}$ is easier to approximate. However, the number 
of $\emme{\q}$s which have to be inverted increases.
Previous works using this approach had to find the best compromise between the length 
of the expansion and the errors introduced by an approximate evaluation of the matrix exponential, 
therefore losing the advantage of an exact expansion.
In order to find a solution to this problem, it is useful to analyze the properties 
of the $\emme{\q}$s in the large $P$ limit. It turns out that matrices with small
$\q$ are much more difficult to handle than those having a higher index. We therefore suggest 
applying different strategies in the two cases.

\subsection{Properties of $\emme{\q}$ matrices}
Let us define the spectral radius of a matrix $\tens{A}$ as the maximum 
modulus of its eigenvalues, $\sigma\left(\tens{A}\right)=\max_i \left|a_i\right|$, and  its 
condition number $\kappa\left(\tens{A}\right)=\sigma\left(\tens{A}\right)/\sigma\left(\tens{A}^{-1}\right)$.
We then introduce the shorthands $\evmax=\sigma\left(\tens{H}\right)$, which is a measure of 
the width of the Hamiltonian's spectrum, and
$\evmin=1/\sigma\left(\tens{H}^{-1}\right)$, which is of the order of the band gap in insulators, 
and tends to zero for metals.
With this notation, the condition number of the Hamiltonian is $\kappa\left(\tens{H}\right)=\evmax/\evmin$. 
In this section, we will obtain the corresponding
quantities for the $\emme{\q}$s. In particular, we will show that $\kappa\left(\emme{\q}\right)$ 
does not depend on $P$ in the large $P$ limit, and demonstrate that the $\emme{\q}$s are always 
better conditioned than the Hamiltonian.

\begin{figure}[tb]
\caption{\label{fig:mq-function} (color online) Plot of 
$\sqrt{1+e^{-x/P}-2e^{-x/2P}\cos\pi \q/P}$, which is equal to $M_{\q}\left(x\right)$ (equation~(\ref{eq:mq-function})) within 
$\mathcal{O}\left(P^{-1}\right)$. 
The dashed line corresponds to the locus of local minima.
}
 \includegraphics{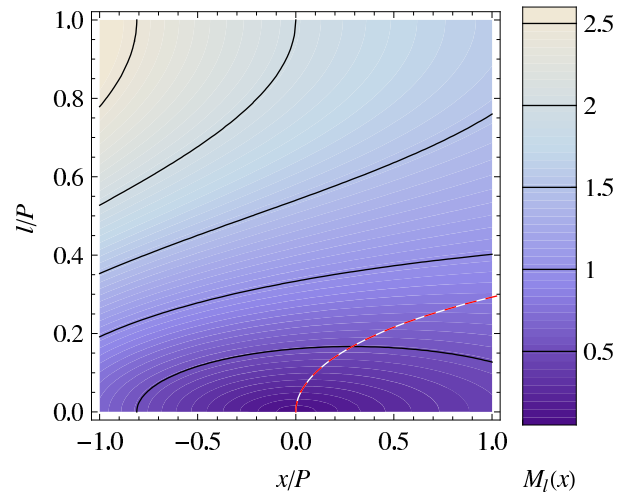}
\end{figure}

We must consider how the spectrum of the Hamiltonian is mapped by the function 
\begin{align}
M_{\q}\left(x\right)=\left|1-e^{\I\pi\left(2\q-1\right)/2P}e^{-x/2P}\right|.
\label{eq:mq-function}
\end{align}
It is readily found that, for any $P$ and $x$, $M_{\q}\left(x\right)$ is a monotonically decreasing function
of $\q$. For fixed $\q$, $M_{\q}\left(\left|x\right|\right)\le M_{\q}\left(-\left|x\right|\right)$  
and the minimum value is $M_{\q}\left(x_{min}\right)=\sin \pi\left(2\q-1\right)/2P$, which is reached 
for $x_{min}=-2P\ln \cos \pi \left(2\q-1\right)/2P$.
From the plot of $M_{\q}\left(x\right)$ (Figure~\ref{fig:mq-function}), it is apparent that the
region which can lead to ill-conditioned matrices is the one with $\q\ll P$ and $x\ll P$, where 
the spectrum of $\emme{\q}$ can contain eigenvalues close to zero.
In this region, an upper bound to the maximum eigenvalue is given by $M_{\q}\left(-\evmax\right)$,
and an estimate of the minimum eigenvalue within $\mathcal{O}\left(1/P\right)$ is $M_{\q}\left(\evmin\right)$.

\begin{figure}[tbh]
\caption{\label{fig:mq-condition} (color online) Condition number of $\emme{\q}$, in the $P\rightarrow\infty$
limit, for a typical value of $\evmax$. Dark (blue) and light (red) series correspond to
the behavior for a metal and for an insulator ($\evmin=20$). Even for a metallic system the condition number
remains finite, and for the insulator it saturates at $\kappa\left(\tens{H}\right)$.
In both cases, $\kappa\left(\emme{\q}\right)$ drops rapidly to one as $\q$ increases.
}
 \includegraphics{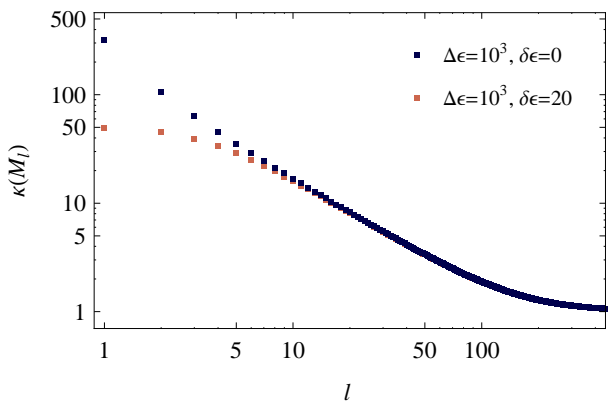}
\end{figure}

The following set of results can easily be proved by series expansion in powers of $1/P$, assuming 
 $\q\ll P$ and $x\ll P$ 
\begin{eqnarray}
\sigma\left(\emme{\q}\right)&=&\frac{1}{2P}\sqrt{\evmax^2+\pi^2\left(2\q-1\right)^2}  
+\mathcal{O}\left(P^{-2}\right) \label{eq:sr-emmeq}\\
\sigma\left(\emme{\q}^{-1}\right)&=&\frac{2P+\evmin/2}{\sqrt{\evmin^2+\pi^2\left(2\q-1\right)^2}} +\mathcal{O}\left(P^{-1}\right) \label{eq:sr-emmeq-inv}\\
\kappa\left(\emme{\q}\right)&=&
\sqrt{\frac{\evmax^2+\pi^2\left(2\q-1\right)^2}
{\evmin^2+\pi^2\left(2\q-1\right)^2}}
+\mathcal{O}\left(P^{-1}\right)\nonumber \\
& \sim &
\kappa\left(\tens{H}\right)
\sqrt{\frac{\evmax^2+\pi^2\left(2\q-1\right)^2}
{\evmax^2+\kappa\left(\tens{H}\right)^2\pi^2\left(2\q-1\right)^2}} \nonumber \\
& \underset{\evmin\rightarrow 0}{\lessapprox} & 1+\frac{\evmax}{\pi\left(2\q-1\right)} 
 \label{eq:cn-emmeq}
\end{eqnarray}
It can be seen from eq.~(\ref{eq:cn-emmeq}) that the condition number $\kappa\left(\emme{\q}\right)$
tends rapidly to one as $\q$ is increased, and is always smaller than $\kappa\left(\tens{H}\right)$ 
(see also Figure~\ref{fig:mq-condition}).
Note that the last inequality in eq.~(\ref{eq:cn-emmeq}), valid for $\evmin\rightarrow 0$, 
shows that $\kappa\left(\emme{\q}\right)$ is bounded also in the metallic case.

\section{A hybrid approach to the expansion \label{sec:hybrid}}
The analysis performed above suggests dealing separately with the few, worst-conditioned 
$\emme{\q}$ matrices having $\q<\bar{\q}$, and with those which have $\kappa\left(\emme{\q}\right)\sim 1$,
for $\q\ge\bar{\q}$. The latter will form the ``tail'' contribution to the density matrix, and will be discussed first.

\subsection{Series expansion for the tail}
\begin{figure}[tb]
\red{
\caption{\label{fig:y-spectrum} (color online) The picture sketches the transformations in the complex plane 
leading from the Hamiltonian spectrum (1) to $\tens{Z}_{\q}$ (2), to $\tens{Z}_{\q}-z'\id$ (3), 
and eventually to $\left(\tens{Z}_{\q}-z'\id\right)/\left(1-z'\right)$ (4). 
The translation (2)$\rightarrow$(3) and the scaling (3)$\rightarrow$(4) depend both on the choice of $z'$.
As described in the text, it is always possible to choose the parameter so as to keep the whole spectrum 
within the unit circle, ensuring convergence of the power series~(\ref{eq:series-yqk}).
}
}
 \includegraphics{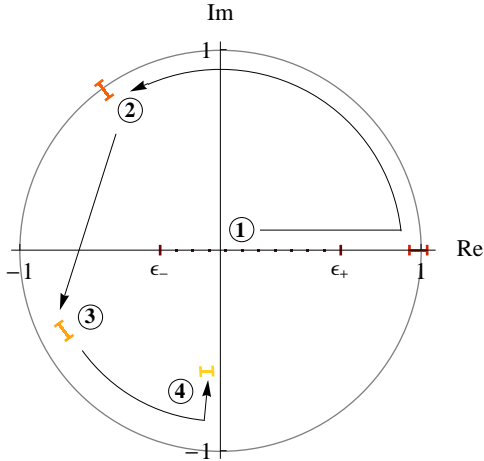}
\end{figure}

In order to obtain a convergent power series for $\emme{\q}^{-1}$,  it is convenient to perform an expansion
around the diagonal matrix $z'\id$, where $z'$ is an arbitrary complex number whose value will be chosen so as to 
accelerate convergence. 
Defining the shorthand $\tens{Z}_{\q}=e^{\I\pi\left(2\q-1\right)/2P}e^{-\tens{H}/2P}$, one has
\begin{eqnarray}
\emme{\q}^{-1}& = &\left(1-z'\right)^{-1}\left[\id-\frac{\tens{Z}_{\q}-z'\id}{1-z'}\right]^{-1}\nonumber\\
& = &\left(1-z'\right)^{-1}\sum_{j=0}^{\infty} \left(\frac{\tens{Z}_{\q}-z'\id}{1-z'}\right)^j.\label{eq:series-yqk}
\end{eqnarray}
The condition for convergence of~(\ref{eq:series-yqk})  is that the whole spectrum of 
$\left(\tens{Z}_{\q}-z'\id\right)/\left(1-z'\right)$ lies within the unit circle in the complex plane.
Moreover, the convergence speed of the expansion will be determined by the eigenvalue which lies farthest 
from the origin (see Figure~\ref{fig:y-spectrum}). 
We refer to appendix~\ref{sec:opt-k} for a detailed analysis of the convergence ratio
\begin{equation}
\chi=\sigma\left(\frac{\tens{Z}_{\q}-ke^{\I\phi_{\q}}\id}{1-ke^{\I\phi_{\q}}}\right), \label{eq:chi-series}
\end{equation}
where we have set $z'=ke^{\I\phi_{\q}}$, defining $\phi_{\q}=\pi\left(2\q-1\right)/2P$, and introducing the 
and complex-valued parameter $k$. There we show that, in the large $P$ limit, one obtains an upper bound to
the convergence ratio, i.e. $\chi=\evmax/\sqrt{\evmax^2+\pi^2\left(2\q-1\right)^2}<1$, provided one 
chooses for the optimal $k$ the analytical estimate 
\begin{equation}
 k=1-\frac{\evmax^2}{2 P \pi\left(2\q-1\right)}.
\label{eq:k-series-guess}
\end{equation}

Having ensured that the series~(\ref{eq:series-yqk}) converges, we can estimate the error made by
truncating the power series after $m_T$ terms, 
\begin{align}
\sigma\left( 
\left(1-z'\right)^{-1}\sum_{j=0}^{m_T} \left(\frac{\tens{Z}_{\q}-z'\id}{1-z'}\right)^j -\emme{\q}^{-1}\right)\le \nonumber \\
\frac{1}{\left|1-z'\right|}\sum_{j=m_T+1}^{\infty}\sigma\left(\frac{\tens{Z}_{\q}-z'\id}{1-z'}\right)^j=
\frac{1}{\left|1-z'\right|}\frac{\chi^{m_T+1}}{1-\chi}. \label{eq:series-error}
\end{align}
In order to achieve a $10^{-D}$ \emph{relative} accuracy on $\emme{\q}^{-1}$, it is necessary to retain at least
\[
 m_{T}\approx \frac{1}{\ln \chi} \left[\ln\left(\frac{1}{\chi}-1\right)-D\ln 10 
+\ln \left(\left|1-z'\right|\sigma\left(\emme{\q}^{-1}\right)\right)\right]
\]
terms. If we use eq.~(\ref{eq:k-series-guess}) and eq.~(\ref{eq:sr-emmeq-inv}), setting $\evmin=0$, 
and taking the large $\evmax$ limit, this estimate takes the simpler form
\begin{equation}
m_{T}\approx 2D  \frac{\evmax^2}{\pi^2\left(2\q-1\right)^2} \ln 10 \label{eq:series-count}
\end{equation}

While the scaling with $\evmax^2$ is not optimal, the dependence on $\q^{-2}$ limits its effects 
to the small-$\q$ terms. These terms can be dealt with effectively with a different 
approach, as we will show below. The influence of the $\evmax^2$ scaling on the overall
operations count will therefore be limited.

Thanks to the chosen $z'$ parametrization the matrix powers entering eq.~(\ref{eq:series-yqk}) 
depend on $\q$ only by a scalar factor,
\[
 \frac{\tens{Z}_{\q}-z'\id}{1-z'}=\frac{e^{\I \phi_{\q}}}{1-ke^{\I \phi_{\q}}}\left(e^{-\tens{H}/2P}-k\id\right)
\]
Therefore, we can compute the expensive powers $\left(e^{-\tens{H}/2P}-k\id\right)^j$ just once,
and obtain any $\emme{\q}^{-1}$ by combining them with the appropriate scalar coefficients.
Furthermore, we often need just the overall contribution to the density matrix
arising from the tail, which reads
\begin{equation}
 \tens{T}_{\bar{\q}}=\sum_{\q=\bar{\q}}^P \emme{\q}^{-1}=
\sum_{j=0}^{m_{T}\left(\bar{\q}\right)}\left(e^{-\frac{\tens{H}}{2P}}-k\id\right)^j
\sum_{\q=\bar{\q}}^P\frac{\left(1-k e^{\I \phi_{\q}}\right)^{-j}}{e^{-\I \phi_{\q}}-k}
\label{eq:series-tail}
\end{equation}
If either $m_T$ or $P$ is very large, computing the scalar coefficients in~(\ref{eq:series-tail})
implies a sizable overhead, which is however independent of the system size, and becomes negligible
for large systems.

\begin{figure}[tb]
\caption{\label{fig:series-k} (color online) The number of terms required to achieve $10^{-3}$ 
relative accuracy in the polynomial expansion of $\emme{\q}^{-1}$ is plotted for a 
Hamiltonian with minimum eigenvalue $-5$, maximum eigenvalue $10$, and for $P=10^4$.
A full line corresponds to results computed keeping $k$ fixed to the $\q=1$ value, while dots 
correspond to the results computed by optimizing $k$ separately for each value of $\q$.
Dark (blue) and light (red) series correspond respectively to the results based on the analytical
estimate~(\ref{eq:k-series-guess}) for $k$, and to the ones obtained by iteratively minimizing 
(\ref{eq:d-series}). Iterative refinement leads to a significant boost in performance. 
In any case, the number of terms computed for $\q=1$ largely exceeds the terms needed to 
compute the contributions for larger $\q$ values, even if $k$ is not optimized on a case-by-case basis.
}
 \includegraphics{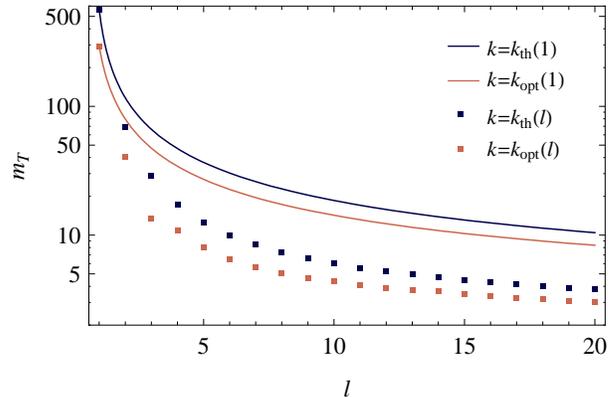}
\end{figure}

In order to assess the accuracy of eq.~(\ref{eq:series-tail}), further analysis is needed. 
If we want to reuse the powers $\left(e^{-\tens{H}/2P}-k\id\right)^j$, we 
must keep $k$ fixed to the value optimized at $\bar{\q}$. Expression~(\ref{eq:series-count}) 
gives the number of terms required to compute $\emme{\q}^{-1}$ with $10^{-D}$ accuracy,
\emph{provided that $k$ is optimized for each $\q$}. 
However, the dependence of $m_T$ on $\q$  offsets the effect of using a non-optimal $k$.
It is easy to show, given the estimate~(\ref{eq:k-series-guess}), that the number of terms
computed for $\bar{\q}$ largely exceeds the number of terms required to compute $\emme{\q}^{-1}$
for any $\q>\bar{\q}$, even if $k$ is kept fixed to the valued optimized for $\bar{\q}$.
Figure~\ref{fig:series-k} shows that this is the case also when $k$ is iteratively optimized
starting from the analytical estimate.

\subsection{Newton inversion in the small-$\q$ region}
To address the inversion of the worst-conditioned terms with $\q<\bar{\q}$, which are 
too expensive to obtain by polynomial expansion, one could resort to one of the 
techniques described in our previous work\cite{kraj-parr05prb,kraj-parr06prb,kraj-parr06prb-2,kraj-parr07prb}. 
In fact, the analysis performed so far can be seen as an improvement to those methods, since we
can evaluate in one shot the contribution from the tail, lowering the number of terms
which must be treated individually, and therefore improving the efficiency.

In this section we will discuss an alternative approach for computing the small-$\q$ $\emme{\q}^{-1}$,
based on a well-established Newton method for matrix inversion.
We give a brief outline of the algorithm and some of its known analytical properties\cite{pan-reif85proc},
and will use them to estimate the number of operations necessary for our purposes.
Given a non-singular, $M\times M$ matrix $\tens{A}$, the iterative procedure
\begin{equation}
\tens{B}_{k+1}=2\tens{B}_k-\tens{B}_k\tens{A}\tens{B}_k  \label{eq:nwt-iter}
\end{equation}
converges to $\tens{A}^{-1}$. Defining $\tens{R}\left(\tens{B}\right)=\id-\tens{B}\tens{A}$,
the condition for convergence is that $\chi=\sigma\left(\tens{R}\left(\tens{B}_0\right)\right)<1$, and the 
error after $k$ iterations is 
\begin{equation}
 \sigma\left(\tens{A}^{-1}-\tens{B}_k\right)=\sigma\left(\tens{B}_0\right)\chi^{(2^k)}\left(1-\chi\right)^{-1},
\label{eq:nwt-residual}
\end{equation}
which corresponds to a number of multiplies (two per iteration) 
\begin{align}
 m_{N}=\frac{2}{\ln 2}
\ln\frac{\ln\left[10^{-D}\left(1-\chi\right)\sigma\left(\tens{A}^{-1}\right)/\sigma\left(\tens{B}_0\right)\right]}{\ln \chi}
\label{eq:newt-niter}
\end{align}
needed to achieve a $10^{-D}$  relative accuracy.

One must then face the problem of finding the approximate inverse $\tens{B}_0$ needed to start the 
iterations~(\ref{eq:nwt-iter}). The authors of Ref.\cite{pan-reif85proc} suggested the simple form 
\begin{equation}
\tens{B}_0=\tens{A}^{\dagger}\left(\left\|\tens{A}\right\|_1 \left\|\tens{A}\right\|_{\infty}\right)^{-1}, \label{eq:nwt-gen-guess}
\end{equation}
where $\left\|\tens{A}\right\|_1=\max_j \sum_{i=1}^{M}\left|A_{ij}\right|$ and  
$\left\|\tens{A}\right\|_{\infty}=\max_i \sum_{j=1}^{M}\left|A_{ij}\right|$.
If one uses eq.~(\ref{eq:nwt-gen-guess}), convergence is guaranteed. Taking as usual the
large $P$ and $\evmax$ limit for a metallic system, one obtains 
$m_{N}\sim \ln \evmax + \ln \left(D\ln 10\right)  + \ln M/\pi^2\left(2\q-1\right)^2 $ 
as an estimate of the operations count to invert $\emme{q}$.
Even if a feeble $M$-dependence has been introduced in the operation count, 
the efficiency is greatly improved if one needs high accuracy or if $\evmax$ is large,
thanks to the exponential convergence rate.

It is however more effective to exploit the simple analytic form for $\emme{\q}^{-1}$ to construct
better initial guesses. For instance, one can use the following relation between $\emme{\q}^{-1}$ 
and $\emme{\q-\delta\q}^{-1}$,
\begin{eqnarray}
 \emme{\q-\delta \q}^{-1}& = & \emme{\q}^{-1}e^{\I\pi\delta \q/P} \left[\id+\emme{\q}^{-1}\left(e^{\I\pi\delta \q/P} -1\right)\right]^{-1} \nonumber\\
& =& e^{\I\pi\delta \q/P} \sum_{j=0}^{\infty}\left(e^{\I\pi\delta \q/P}-1\right)^j \emme{\q}^{-(j+1)} \label{eq:newt-extra}
\end{eqnarray}
to estimate a guess for $\emme{\q-\delta\q}^{-1}$ starting from an already-computed inverse.
The series~(\ref{eq:newt-extra}) converges provided that $\left|e^{\I\pi\delta \q/P}-1\right|\sigma\left(\emme{\q}^{-1}\right)<1$.
In the $P\rightarrow\infty$ limit this amounts to the condition $\delta \q<\q -1/2$. In theory, all the terms up to $\q=1$
could be computed inserting any $\emme{\q}^{-1}$  into eq.~(\ref{eq:newt-extra}).
In practice, computing powers of $\emme{\q}^{-1}$ is not advisable if we aim at linear scaling, since the $\emme{\q}^{-1}$s 
and their powers tend to be much fuller than the Hamiltonian, and the asymptotic convergence rate of 
eq.~(\ref{eq:newt-extra}) is worse than the one for the iterative inversion. 
In any case, the lowest-order approximation is already much more effective than the universal 
guess described in Ref.~\cite{pan-reif85proc}.
One finds that the convergence ratio for the computation of $\emme{\q-1}^{-1}$, using 
the low-order extrapolation $e^{\I\pi/P}\emme{\q}^{-1}$, is 
$\chi\sim 2\pi/\sqrt{\evmin^2+\pi^2\left(2\q-1\right)^2}$, 
leading to an estimate of the number of the operations count
\begin{equation}
 m_{N}=\frac{2}{\ln 2}\ln\frac{D\ln 10 -2\ln \pi}{\ln q} +\mathcal{O}\left(1/q\right) \label{eq:newt-count-extra}
\end{equation}
This estimate is independent of $\evmin$ because we considered the worst-case scenario where
the system is metallic. It is also independent of $M$ and - most importantly - of
$\evmax$.
In practice, one starts from $\emme{\bar{\q}}^{-1}$ obtained from the polynomial expansion, then computes
$\emme{\bar{\q}-1}^{-1}$, using $e^{\I\pi/P}\emme{\bar{\q}}^{-1}$ as the initial guess, and continues 
stepwise, obtaining the initial estimate for iterative inversion of $\emme{\bar{\q}-2}$ from the 
previously computed $\emme{\bar{\q}-1}^{-1}$, and so on. Alternatively, the first inverse matrix
$\emme{\bar{\q}-1}^{-1}$ can be computed starting from the simple guess~(\ref{eq:nwt-gen-guess}).
Efficient higher-order extrapolations will be discussed in appendix~\ref{sec:ho-extra}.

\subsection{Overall operation count}

\begin{figure}[tb]
\caption{\label{fig:tot-mult} (color online) Total number of matrix-matrix multiplications required
to obtain the density matrix, combining series expansion and Newton inversion methods, on a 
$\log$-$\log$ plot. Light (red) and dark (blue) lines correspond to $10^{-5}$ and $10^{-8}$ target
accuracy respectively. Full (a), dashed (b) and dotted (c) lines correspond respectively to
the number of operations estimated using the general-purpose initial estimate, using a zero$^{th}$-order
extrapolation guess and using extrapolation together with fast polynomial evaluation in the tail
region. Grid lines mark the slope expected for a linear dependence between $\evmax$ in units 
of $k_B T$ and the overall operations count $m_{T}$.
}
 \includegraphics{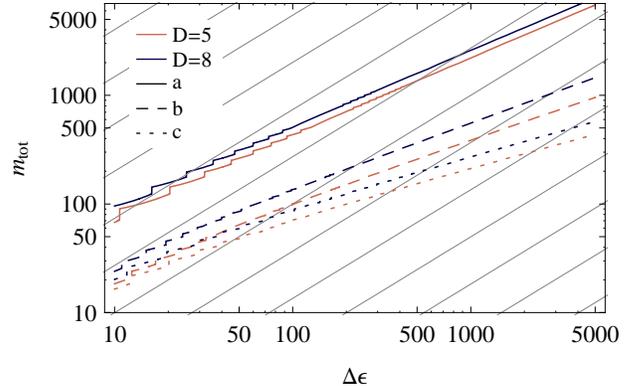}
\end{figure}

In the previous section we obtained (equations~(\ref{eq:series-count}) and~(\ref{eq:newt-niter}))
an upper bound estimate of the number of matrix-matrix multiplications needed in order to obtain the tail
contribution up to $\bar{\q}$, and to invert a single $\emme{\q}$ using an iterative Newton method.
The optimal value for $\bar{\q}$ is obtained when the incremental cost of including an extra term in the
tail contribution $\tens{T}_{\q}$ (cfr. eq.~\ref{eq:series-tail}) becomes larger than the cost of a single 
iterative inversion, i.e. when
\begin{equation}
m_{T}\left(\bar{\q}\right)-m_{T}\left(\bar{\q}+1\right)\ge  m_{T}\left(\bar{\q}\right).
\label{eq:best-q}
\end{equation}
The overall number of multiplications is then
\begin{equation}
m_{tot}=m_{T}\left(\bar{\q}\right)+ \sum^{\bar{\q}-1}_{\q=1} m_{N}\left(\q \right).
\label{eq:tot-mult}
\end{equation}

 In figure~\ref{fig:tot-mult} we plot the overall operations count obtained by using our theoretical estimates
 for $m_{T}$ and $m_N$.
A dramatic improvement is obtained when we use $e^{\I\pi/P}\emme{\q}^{-1}$ as the initial guess 
for the inversion of $\emme{\q-1}$.
We can think of the extrapolated guess as an almost optimal preconditioner 
and are considering how this could be exploited in different inversion schemes as well.
 It is worth noting that - despite the fact that the tail contribution requires
a number of multiplies scaling quadratically with $\evmax$ - the overall scaling is significantly
sublinear. \red{Comparing our results (figure~\ref{fig:tot-mult}b) with the multiplication count for 
standard Chebyshev polynomials expansion, as given by Ref.\cite{baer-head97jcp}, 
our method becomes beneficial by $\evmax \sim 20$ - 
the break-even point getting lower as the target accuracy $D$ is increased.}
Fast polynomial summation methods\cite{lian+03jcp,lian+04jcomp,vanl79ieee} can be used to compute both
$\emme{\bar{\q}}^{-1}$ and $\tens{T}_{\bar{\q}}$. This reduces the number of multiplies from $m_T$ to
$3\sqrt{m_T}$, however at the cost of storing an extra $\sqrt{m_T}$ matrices.
Combining these fast summation techniques with iterative inversion
further lowers the operations count, leading to a scaling slightly better than $\sqrt{\evmax}$
(figure~\ref{fig:tot-mult}c). 
\red{In this case, however, the prefactor of our method is larger, so that the break-even point,
when comparing with Ref.\cite{lian+03jcp,lian+04jcomp,vanl79ieee}, is shifted towards higher 
accuracy and large $\evmax$. We are currently investigating the possibility of 
applying an alternative expansion of the tail contribution, which should make both our $\evmax$ scaling
and the prefactor highly competitive.}

\section{A test case \label{sec:test-case}}
So far we have estimated the accuracy of the computation of each $\emme{\q}^{-1}$ term using 
 $\varepsilon\left(\tens{\tilde{M}}_{\q}^{-1}\right)=
\sigma\left(\emme{\q}^{-1}-\tens{\tilde{M}}_{\q}^{-1}\right)/\sigma\left(\emme{\q}^{-1}\right)$
as a measure of the error affecting the estimate $\tens{\tilde{M}}_{\q}^{-1}$.
However, the quantity we are more interested in is the band structure energy
$E=\Tr \left[\tens{\boldsymbol{\rho}}\tens{H}\right]$. A theoretical estimation of the error on $E$ requires 
several assumptions on the distribution of errors over the different eigenvalues of the Hamiltonian, and
the different $\q$ terms, and we have not attempted it here.
We have instead tested our method against a real system, selecting the self-consistent DFT Hamiltonian
matrix of a 128-atom sample of the metallic fcc phase of $\mathrm{LiAl}$, as computed by the 
CP2K\cite{vand-krac05cpc,CP2K} package\footnote{We used GTH pseudopotentials\cite{goed+96rb,krac05tca,hart+96prb}, 
with PBE\cite{perd+96prl} exchange-correlation
functional, a double-$\zeta$ basis with one additional set of polarization functions, for a total of 
1728 basis functions. $\mathrm{LiAl}$ in the fcc phase is a metal. Since we are computing the Hamiltonian at the 
$\Gamma$ point only, the spectrum has six half-occupied degenerate states at the zero-temperature Fermi energy.
In the low-temperature limit $\evmin\rightarrow 0$, which makes this system particularly challenging.}.
The orthogonal Hamiltonian matrix is obtained by multiplying the non-orthogonal one with the
inverse square root of the overlap matrix\cite{lowd50jcp}. We the computed with
standard diagonalization techniques the chemical potential and the exact band-structure energy
for different electronic temperatures. We also obtained the bounds of the spectrum of $\tens{H}$ 
 ($\epsilon_+=121.15$~eV and $\epsilon_-=-42.65$~eV), which are needed in eq.~(\ref{eq:d-series}) and 
could in principle be computed in linear scaling with the Lanczos method, or easily estimated 
by Gershgorin's circle theorem\cite{pals-mano98prb} or any matrix norm.

\begin{figure}[tb]
\caption{\label{fig:test-oc-accu} (color online) Number of matrix-matrix multiplications used to
achieve a given error on the band structure energy, for different electronic temperatures. Details of the 
system are given in the text. The data points for every temperature, from left to right, 
correspond to $10^{-2}$, $10^{-3}$, $10^{-5}$, and $10^{-7}$ target accuracy. 
}
 \includegraphics{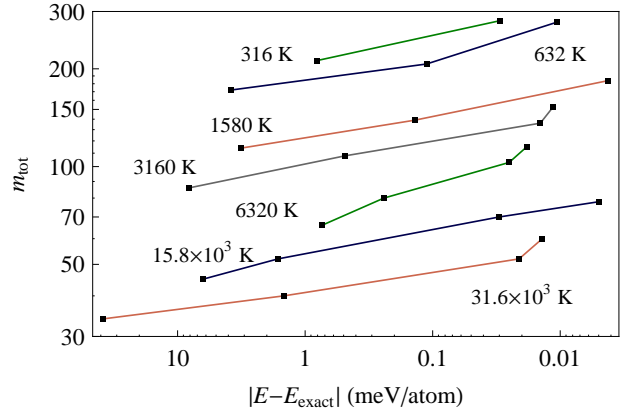}
\end{figure}

We then applied our algorithm to the orthogonalized Hamiltonian, using fast polynomial summation to compute the tail 
and using first-order extrapolation in the Newton region, with a history vector 
containing the last two matrices (cfr. eq.~(\ref{eq:extra-first-order})).
Slight improvements in the operations count could be obtained by hand-tuning $\bar{\q}$,
but we just used the automatic procedure based on our theoretical estimates, as described in the previous section.
In Figure~\ref{fig:test-oc-accu} we plot the number of multiplications performed versus the 
resulting error on the energy. Since we can use a large value of $P$, $e^{-\tens{H}/2P}$ can be computed
with only a few matrix-matrix multiplies, which have not been included in the operations count.

\begin{figure}[tb]
\caption{\label{fig:test-opcount} (color online) 
The number of matrix-matrix multiplies performed to compute the density matrix for the $\mathrm{LiAl}$ test
case, for different electronic temperatures and target accuracies, plotted on a $\log - \log$ scale,
together with guidelines corresponding to a $m\propto \sqrt{\evmax}$  scaling.
}
 \includegraphics{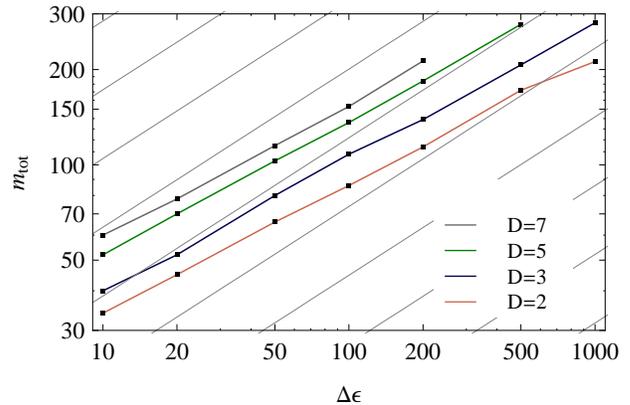}
\end{figure}

For a given target accuracy, the operations count scales better than $\sqrt{\beta}$ (Figure~\ref{fig:test-opcount}). 
We also observe that the accuracy of the energy is 
much better than the relative accuracy guaranteed by the theoretical estimates. Consider for example that, 
by requiring a relative ``spectral radius accuracy'' better than $10^{-2}$ 
(first data points in Figure~\ref{fig:test-oc-accu}) we obtain a relative error on the energy
of the order of $10^{-4}$ (the total energy is $\sim -5$~keV). This is mainly due to the fact that the 
error in the energy is second order with respect to the error in the density matrix.
However, we observe that also the error in the full density matrix, computed as the spectral radius of the difference
with the result obtained with diagonalization, is in general almost one order of magnitude smaller
than the required accuracy. This result is probably due to a combination of effects: firstly,
we use worst-case estimates, so that the accuracy of the individual terms is necessarily higher
than the assumed one. Moreover, the errors affecting different $\q$ terms might partially cancel each other out,
and many of the contributions in the Newton region are computed with an accuracy much higher than
requested, due to the exponential convergence.
The accuracy improves very quickly as the number of operations increases until, for errors around $0.01$~meV/atom,
numerical issues come into play and prevent further refinement, which is anyway hardly necessary for
most applications.

\red{
 Most of the observables relevant to electronic structure calculations, such as forces 
and electronic density, are readily evaluated by expressions of the form 
$\left<A\right>=\Tr \left[\tens{\boldsymbol{\rho}}\tens A\right]$.
Since the matrix $\tens{A}$ obeys the same sparsity as the Hamiltonian
$\left<A\right>$ depends only on a small subset of the nonzero elements of the density matrix.
We are currently investigating whether it is possible to compute the expectation value directly, 
without evaluationg non-relevant elements of $\tens{\boldsymbol{\rho}}$, 
which would further improve the efficiency.}

\section{Conclusion}
We have performed a detailed study of a recently-proposed form for Fermi operator expansion.
The properties of this expansion allow features of the expansion in polynomial and 
rational functions to be combined, and by optimizing the mixture we can 
have the best of both worlds.
In this way, we circumvent the tradeoff between the number of terms and the accuracy
of the expansion, which was needed by prior implementations of this expansion of the Fermi operator.
Moreover, sub-linear scaling of the matrix-matrix multiplications count with respect to the Hamiltonian 
range is achieved, making the method particularly attractive for low-temperature and high-accuracy applications.
However, there is still room for improvement. In particular, work is in progress  in the direction of a 
better polynomial expansion in the tail region. We are also considering applying the method to 
molecular dynamics. In this case one could use the $\emme{\q}^{-1}$s stored from the previous step as
a starting point for iterative minimization. In this way, the computation of the different $\q$-channels
can be made independent, adding a layer of parallelism on top of the parallel matrix-matrix multiply.
Formal analogies between our expansion and Trotter factorization entering path integral techniques 
suggest that some of the ideas presented here might be useful to tackle that problem as well.
In order to achieve linear scaling, attention should be paid to the issue of matrix truncation, 
since here we have dealt only with matrix-matrix operations counts.
 Preliminary results show that in this respect there are no significant differences
from standard expansion methods, as the minimum sparsity of the terms taken into
account is basically the same as the sparsity of the whole density matrix, which is dictated by the
physics of the system. 
The detailed analysis we have performed in this work has allowed us to obtain significant improvements over
the previous applications of this decomposition of the Fermi operator, and lays solid foundations 
for further progress.

\section{Acknowledgments}
The generous allocation of computer time by the Swiss National Supercomputing Center (CSCS) and
technical assistance from Neil Stringfellow is kindly acknowledged.
We would also like to thank Giovanni Bussi and Paolo Elvati for fruitful discussion.

\appendix
\section{Optimal parameter for series expansion\label{sec:opt-k}}
We show how the value of $k$ in eq.~(\ref{eq:chi-series}) can be 
optimized to obtain faster convergence of the polynomial expansion.
Expressions involved are quite lengthy, so we introduce several shorthands.
Let $\epsilon_{\pm}$ be the bounds of the Hamiltonian spectrum. 
We parametrize $k$ as $k=\left(1+e^{\I\theta}r/P\right)$,
define $e^{\I\pi\left(2\q-1\right)/2P}=\left(v_{\q}+\I w_{\q}\right)$ and 
$s_{\pm}=e^{-\epsilon_{\pm}/2P}$. 
The square modulus of the extrema of the transformed Hamiltonian 
spectrum  (see Figure~\ref{fig:y-spectrum}) is
\begin{equation}
d_{\pm}^2=
\frac{r^2+P^2\left(s_{\pm}-1\right)^2-2 P r \left(s_{\pm}-1\right)\cos\theta}
{r^2+2 P^2\left(1-v_{\q}\right) + 2 P r \left(\cos\theta\left(1-v_{\q}\right) +w_{\q}\sin\theta\right)}
\label{eq:d-series}
\end{equation}
and the convergence ratio is $\chi=\max\left(d_+, d_-\right)$.
One can obtain an analytical estimate for $k$, and an upper bound for $\chi$, by taking 
the $P\rightarrow\infty$ limit, and making the simplifying assumption 
$\left|\epsilon_{\pm}\right|=\evmax$.
 This implies $\theta=-\pi/2$ and leads to the estimate~(\ref{eq:k-series-guess}), 
which can be further improved by minimizing numerically~(\ref{eq:d-series}) 
with respect to $\theta$ and $r$.

\section{High-order initial guess for iterative inversion\label{sec:ho-extra}}
One can derive expressions for high-order extrapolation of inverse $\emme{\q}$ 
matrices from equation~(\ref{eq:newt-extra}), writing them as a linear combination 
of already-computed inverses. We will sketch the procedure by deriving the expression
for the first-order extrapolation of $\emme{\q-1}^{-1}$, using only $\emme{\q}^{-1}$ and 
$\emme{\q+1}^{-1}$, which is then easily extended to higher orders.
Let $c^{(n)}_j=e^{-n\I\pi/P}\left(e^{-n\I\pi/P}-1\right)^{j-1}$. One can write
the first-order extrapolations for the new inverse and for the already-computed one, as
a function of powers of $\emme{\q}^{-1}$:
\begin{eqnarray*}
 \emme{\q-1}^{-1}=c^{(-1)}_1 \emme{\q}^{-1} + c^{(-1)}_2 \emme{\q}^{-2} \nonumber\\
 \emme{\q+1}^{-1}=c^{(1)}_1 \emme{\q}^{-1}  + c^{(1)}_2  \emme{\q}^{-2} \nonumber\\
\end{eqnarray*}
This linear system can be solved for $\emme{\q-1}^{-1}$ and $\emme{\q}^{-2}$, obtaining
\begin{equation}
 \emme{\q-1}^{-1}\sim \emme{\q}^{-1}\left(e^{\I\pi/P}+e^{2\I\pi/P}\right)-\emme{\q+1}^{-1}e^{3\I\pi/P}.
\label{eq:extra-first-order}
\end{equation}
For higher orders one simply inserts into the system more constraints, corresponding 
to ``older'' inverse matrices, and writes the extrapolation including higher powers of 
$\emme{\q}^{-1}$. The system is then solved in terms of these powers, eventually 
finding the coefficients for the estimate of the new inverse as a linear combination of
the older ones.

\end{document}